\documentclass[fleqn,10pt]{wlscirep}
\usepackage[normalem]{ulem} % either use this (simple) or

\title{Ice-Templated W-Cu Composites with High Anisotropy}

\author[1,2,+,$\dag$]{Andr\'{e} R\"othlisberger}
\author[1,$\ddag$]{Sandra H\"aberli}
\author[1,+,*]{Henning Galinski}
\author[3]{David C. Dunand}
\author[4,*]{Ralph Spolenak}
\affil[1]{Laboratory for Nanometallurgy, Department of Materials, ETH Zurich, Vladimir-Prelog-Weg 1-5/10, CH-8093 Z\"urich, Switzerland}
\affil[2]{Mechanical Integrity of Energy Systems, Swiss Federal Laboratories for Materials Science and Technology, EMPA, CH-8600 D\"ubendorf, Switzerland}
\affil[3]{Department of Materials Science and Engineering, Northwestern University, Evanston, IL 60208, USA}

\affil[*]{henning.galinski@mat.ethz.ch and ralph.spolenak@mat.ethz.ch}

\affil[+]{these authors contributed equally to this work}
\affil[$^\dag$]{currently at BIOTRONIK, Switzerland}
\affil[$^\ddag$]{currently at ABB, Switzerland}
\begin{abstract}
Controlling anisotropy in self-assembled structures enables engineering of materials with highly directional response. Here, we harness the anisotropic growth of ice walls in a thermal gradient to assemble an anisotropic refractory metal structure, which is then infiltrated with Cu to make a composite. Using experiments and simulations, we demonstrate on the specific example of tungsten-copper composites the effect of anisotropy on the electrical and mechanical properties. The results are compared to isotropic tungsten-copper composites fabricated by standard powder metallurgical methods. Our results have the potential to fuel the development of more efficient materials, used in electrical power grids and solar-thermal energy conversion systems. The method presented here can be used with a variety of refractory metals and ceramics, which fosters the opportunity to design and functionalize a vast class of new anisotropic load-bearing hybrid metal composites with highly directional properties.
\end{abstract}

\begin{document}

\flushbottom
\maketitle
\thispagestyle{empty}

%%%%%%%%%%%%%%% Introduction

\section*{Introduction}

The number of materials performing well in extreme environments, such as high temperatures above $2000^\circ$C, high voltage or hazardous radiation fields, is limited and typically comprises high-melting materials, for example oxides~\cite{Ullmann2000} and metal alloys based on niobium (Nb), molybdenum (Mo), tantalum (Ta), tungsten (W) and rhenium (Re).  These metals, commonly dubbed ``refractory metals'' and their alloys find wide application in nuclear reactors~\cite{Zhu2013107,Lassner1999}, turbines~\cite{Lassner1999}, space-crafts~\cite{Lassner1999}, thermal emitters~\cite{Ilic2016}, heat spreaders~\cite{Lassner1999,Zweben1992} and circuit breakers~\cite{Gessinger1} in power grids. Moreover, refractory metals are attractive candidates for plasmonic~\cite{Zhoue1501227,Guler263} or photonic~\cite{Rinnerbauer2012,Rinnerbauer2014,Celanovic2008,Arpin2011} materials serving in solar-thermal energy conversion systems. Nevertheless, there are still substantial limitations regarding the fabrication and processing of new refractory metal-composites. The large melting point fabrication requires high temperature annealing making the design of new materials extremely challenging.\\
Of particular interest is the assembly of functional composites with anisotropic phase orientations~\cite{wicklein2015thermally,Sander2016,LeFerrand2016,ADFM:ADFM201300340,doi:10.1021/acsnano.6b07932,Munch1516} which would allow for a highly directional materials response, e.g. anisotropic thermal or electrical conductivity. %But interesting templating techniques such as magnetic templating~\cite{Sander2016,LeFerrand2016} do not apply and 
The fabrication of refractory metal based composites is typically restricted to powder metallurgical techniques. These powder-based methods have significant shortcomings because the materials produced are limited to isotropic composite structures with limited access to their properties and unavoidable residual porosity.\\  
Here, we report on a different approach to create refractory metal-based composites utilizing ice-templating~\cite{Deville2006,Deville2017Book} in a thermal gradient to assemble an anisotropic refractory metal scaffold which is then infiltrated by a second liquid phase. Ice-templating or freeze-casting is a remarkable process, using the anisotropic growth kinetics of ice during freezing to assemble suspended particles in a lamellar microstructure. The thickness of lamellae within the composite can be controlled by altering the velocity of the freezing front, while their volume fractions is given by the particle concentration in the suspension. This method has been successfully applied to fabricate various of complex composites, including synthetic nacre~\cite{Mao107}, bio-materials~\cite{Wegst2099}, porous ceramics~\cite{Deville2008}, ceramic-polymer composites~\cite{Munch1516} and metals~\cite{PARK2017435,Roethlisberger_2016,LI2011146,met5041821,Park2016,YOOK20091502,YOOK20091502}.
%%%%%%%%%%%%%%%%%%%%%
\begin{figure*}[t!]
\begin{center}
\includegraphics[width=0.95\textwidth]{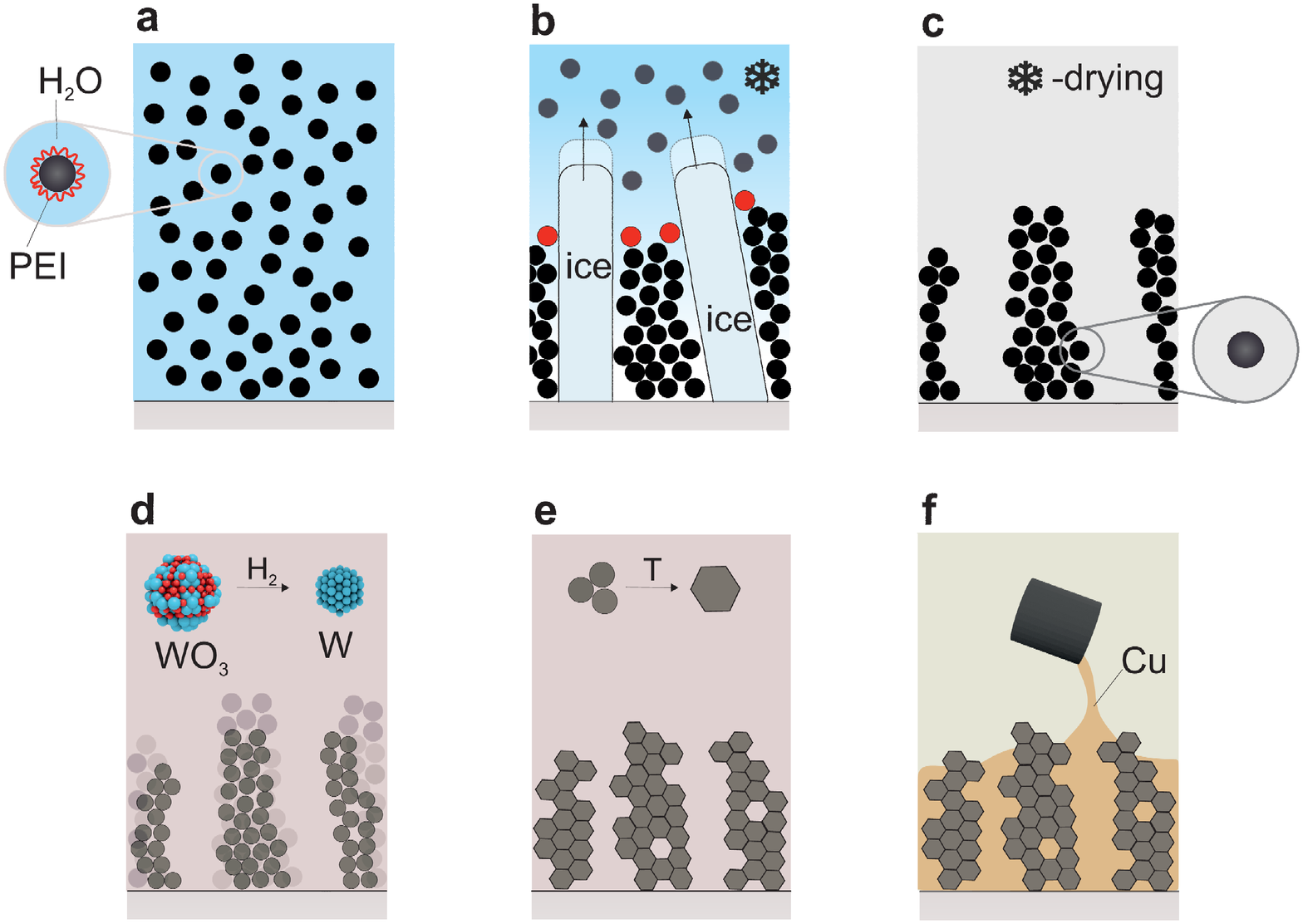}
\caption{\textbf{Schematic of the ice-templating process to create the anisotropic tungsten lamellar scaffold for melt-infiltration.} The process steps involve (a) slurry formation by dispersing the ceramic powder (WO$_3$) in water using PEI as surfactant, (b) formation of an anisotropic composite by ice-templating, particles (red) are captured between the ice-walls, i.e. lamellar dendrites (c) sublimation of ice to yield a loosely-bound WO$_3$ scaffold with directional porosity, (d) reduction of the (WO$_3$) scaffold yielding a metallic tungsten scaffold and (e) sintering of the porous tungsten. Significant volume shrinkage is associated with densification upon reduction (ceramic to metal) and sintering (closing of interparticle porosity). (f) Melt infiltration of the tungsten scaffold with liquid copper followed by soldification.}
\label{fig:1}
\end{center}
\end{figure*}
%%%%%%%%%%%%%%%%%%%%%
We have selected W-Cu composites as specific example, as W-Cu composites offer unique heat- and electrical transport properties paired with excellent mechanical strength at ambient and elevated temperatures, due to the outstanding thermal and electrical conductivity of copper combined with the high strength of tungsten. Historically, W-Cu composites proved to be an ideal testbed to understand fibre-reinforcement in composites~\cite{KELLY1965}. These findings are still seminal for the understanding and functionalization of bio-inspired materials~\cite{Libonati2017}, such as platelet reinforced polymers~\cite{Bonderer1069,Behr2015} or artificial nacre~\cite{Morits2017,Finnemore2012}. W-Cu composites excel in applications where improved heat and current transport properties under extreme mechanical or thermal conditions are needed~\cite{Kim2001,Ibrahim2009} including electrodes for electrical discharge machining~\cite{Wang2009,Marafona1}, arcing contacts in high voltage circuit breakers~\cite{Gessinger1,Vuellers2015213}, fusion energy applications~\cite{Lassner1999,Zweben1992} or heat spreaders with low coefficient of thermal expansion~\cite{Lassner1999,Chung2012}.\\
\\
We fabricated anisotropic W-Cu composites using a three-step process illustrated in Figure~\ref{fig:1}(a)-(f), starting with ice-templating a tungsten oxide (WO$_3$) precursor suspended in water (Figure~\ref{fig:1}(b)-(c)). Directional solidification is used to direct the assembly of a highly directional frozen green body, that is reduced and sintered to a metal scaffold by annealing (Figure~\ref{fig:1}(c)-(e)). In a final step, this lamellar metallic scaffold is infiltrated by liquid copper and solidified, as shown in Figure~\ref{fig:1}(f). Detailed information on the synthesis and processing of the tungsten skeleton is given in a previous work~\cite{Roethlisberger_2016}.\\
%%%%%%%%%%%%%%%%%%%%%
\begin{figure*}[t!]
\begin{center}
\includegraphics[width=0.98\textwidth]{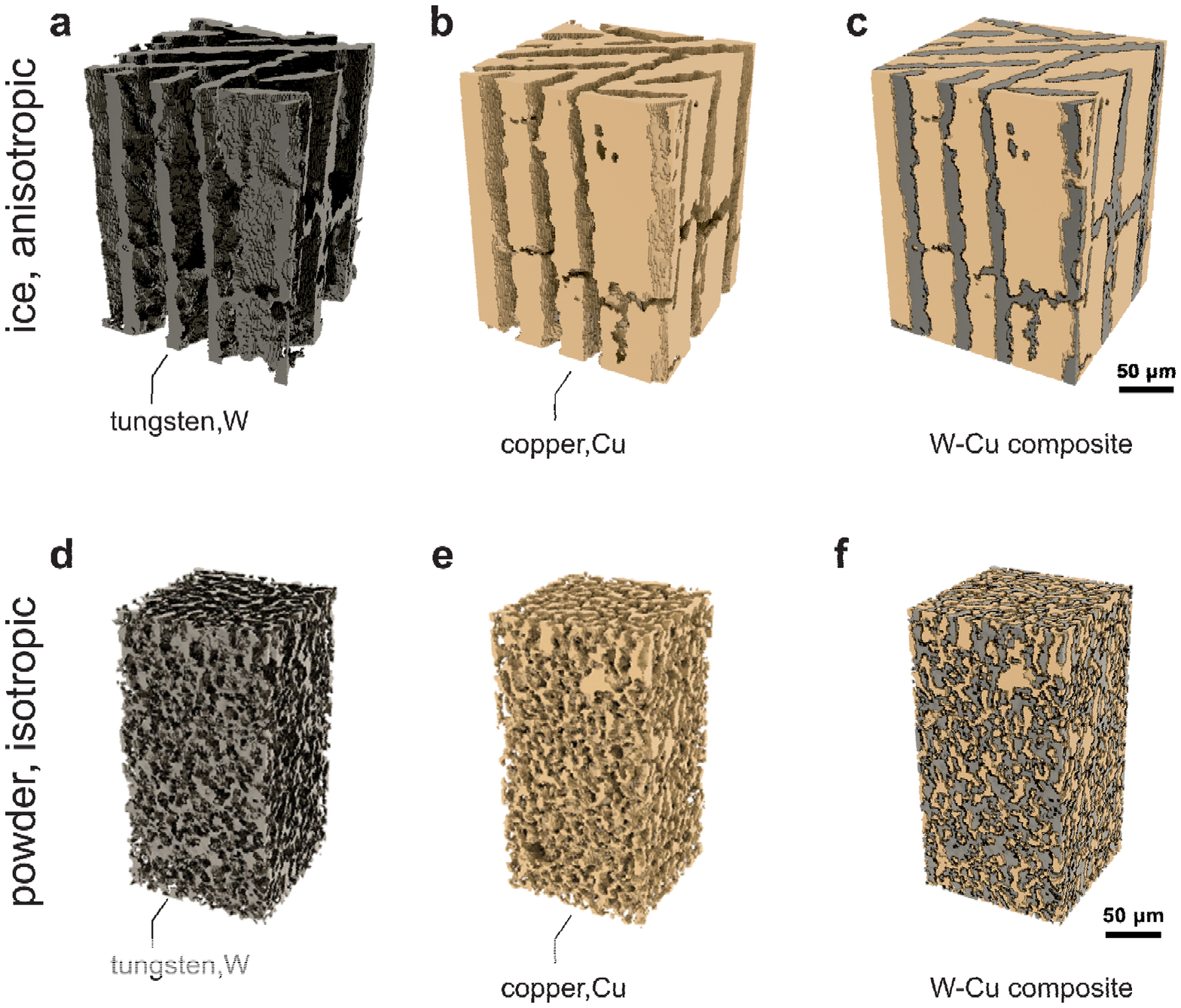}
\caption{\textbf{Composite Architecture.} Three-dimensional X-ray computed tomographic reconstructions of W-Cu composites. (a-c) Show the Cu (a) and W (b) phases of an anisotropic lamellar composite produced by freeze-casting and reduction of a WO$_3$ precursor, sintering and Cu infiltration; the volume fraction of W is $57\%$ W/Cu. Panels (d-f) show the different phases of an isotropic W-Cu composite (65 vol.\% W) produced by partial sintering  of W powders followed by Cu infiltration. For both reconstructions the voxel size is 1.3 $\mu$m.}
\label{fig:2}
\end{center}
\end{figure*}
%%%%%%%%%%%%%%%%%%%%%
To investigate the role of anisotropy in this composite, the microstructural, electrical and mechanical properties of the ice-templated W-Cu composite were measured and compared to isotropic W-Cu composites fabricated by powder metallurgy. High-resolution x-ray computed tomography (XCT) provided three-dimensional information about the microstructure and anisotropy, four-point probe resistivity measurement gave insight on the impact of anisotropy on the electrical properties and uniaxial compression test were used to analyze the mechanical response of the composites. The results are discussed in the following.
%%%%%%%%%%%%%%% Results and discussion
\section*{Results and Discussion}
\subsection*{Composite Architecture}

To image the microstructure of the ice-templated and powder metallurgy based W-Cu composites, we resort to high resolution x-ray computed tomography (XCT). While allowing for a sub-micron resolution, XCT is a widely used, non-destructive technique to characterize complex materials and composites~\cite{pietsch2017x,Withers2014,pietsch2016combining,maire2012x}. Figure \ref{fig:2} shows reconstructed three-dimensional images, including the composites individual phases, of the two analyzed W-Cu composites. The composite fabricated by ice-templating shown in Fig.~\ref{fig:2}(a-c) exhibits a lamellar architecture with a high structural anisotropy. The tungsten skeleton has been fully infiltrated by copper, resulting in ordered composite with a volume fraction of $57\%$ for W and $43\%$ for Cu.\\
Image analysis of radial and longitudinal cross-section (see Supplementary Information, SFig.2) revealed the presence of a 4.0 \% porosity within the ice-templated composites. Two sources for this porosity are identified: (i) closed porosity arising from the tungsten foam synthesis during freeze-casting, which was measured for this particular foam by He-pycnometry to be 5 vol.\% and (ii) incomplete wetting of the tungsten walls by the copper melt. To quantify the anisotropy, i.e. determine the spatial periodicity and orientation of the lamellae within the composite, we used three-dimensional fast Fourier transform (3D FFT), detailed information is given in the Supplementary Information(SFig.1). Two sets of lamellae rotated 40$^\circ$ relative to each other with a mean spatial frequency of $34(2)$~$\mu$m have been identified. Quite interestingly, the height of the ice-walls is generally not constraint in growth direction as long as the longitudinal thermal gradient is maintained. Accordingly, ice-templated composites can be fabricated, where the aspect ratio, i.e. height/width of each lamella is much larger than achieved here. This situation resembles platelet reinforced composites~\cite{Bonderer1069,Libanori2016}, where the aspect ratio of the platelets critically impacts the mechanical properties.\\
Conversely, when powder metallurgical methods are used, the composite as shown in Fig.~\ref{fig:2}(d-f) is highly isotropic and characterized by two intermingled percolating phases. The structural analysis by 3D FFT (SFig.1 Supplementary Information) reflects this result showing no directionality and reveals a spacing between the Cu and W phases of $9$~$\mu$m.
In both case, the spatial frequency of the composite is defined during the tungsten scaffold formation. For the powder metallurgical process, the powder size as well as the pressure and temperature during hot pressing determine the size, fraction and tortuousity of the porosity and thus the overall composite morphology after melt-infiltration with copper.
\\
In the case of ice-templating, the scaffold architecture (W lamella thickness, channel width and orientation ) is defined by the fraction of powder in the slurry : the W lamella thickness increased from 28 to 49 $\mu$m with increasing the slurry particle fraction from 30 to 35 vol.\% (Tab.1). Also important is the solidification velocity, which controls the dendrite spacing and the extent of pushing and accumulation of the ceramic nanoparticles between the growing ice dendrites \cite{Deville2008,Zhang2005, JACE01094,Deville2006}.  Finally, the sintering temperature and time (and the use of sintering aid) control the porosity within the walls, as well as the extent of shrinkage of the scaffold and thus the final W wall and channel width.  

%In case of ice-templating, the scaffold architecture is defined by the pushing and accumulation of the ceramic nanoparticles between the growing ice dendrites upon solidification of the slurry, which in turn is dependent on the ice growth kinetics~\cite{Deville2008,Zhang2005, JACE01094,Deville2006}. Ice solidification as well as particle pushing and packing in the interdendritic space, define the W lamella thickness and the open pore channel size and orientation in the scaffolds. In addition, the lamella thickness of W-foams produced by ice-templating increased slightly with the slurry particle loads (30 and 35 vol.\% respectively) from was 28 $\mu$m to and 49 $\mu$m (Tab.1), respectively. This difference arises from the higher solid fraction in the slurry needed to achieve the high tungsten content foam in the latter case.}

\subsection*{Electrical Properties}

%%%%%%%%%%%%%
\begin{figure*}[h!]
\begin{center}
\includegraphics[width=0.98\textwidth]{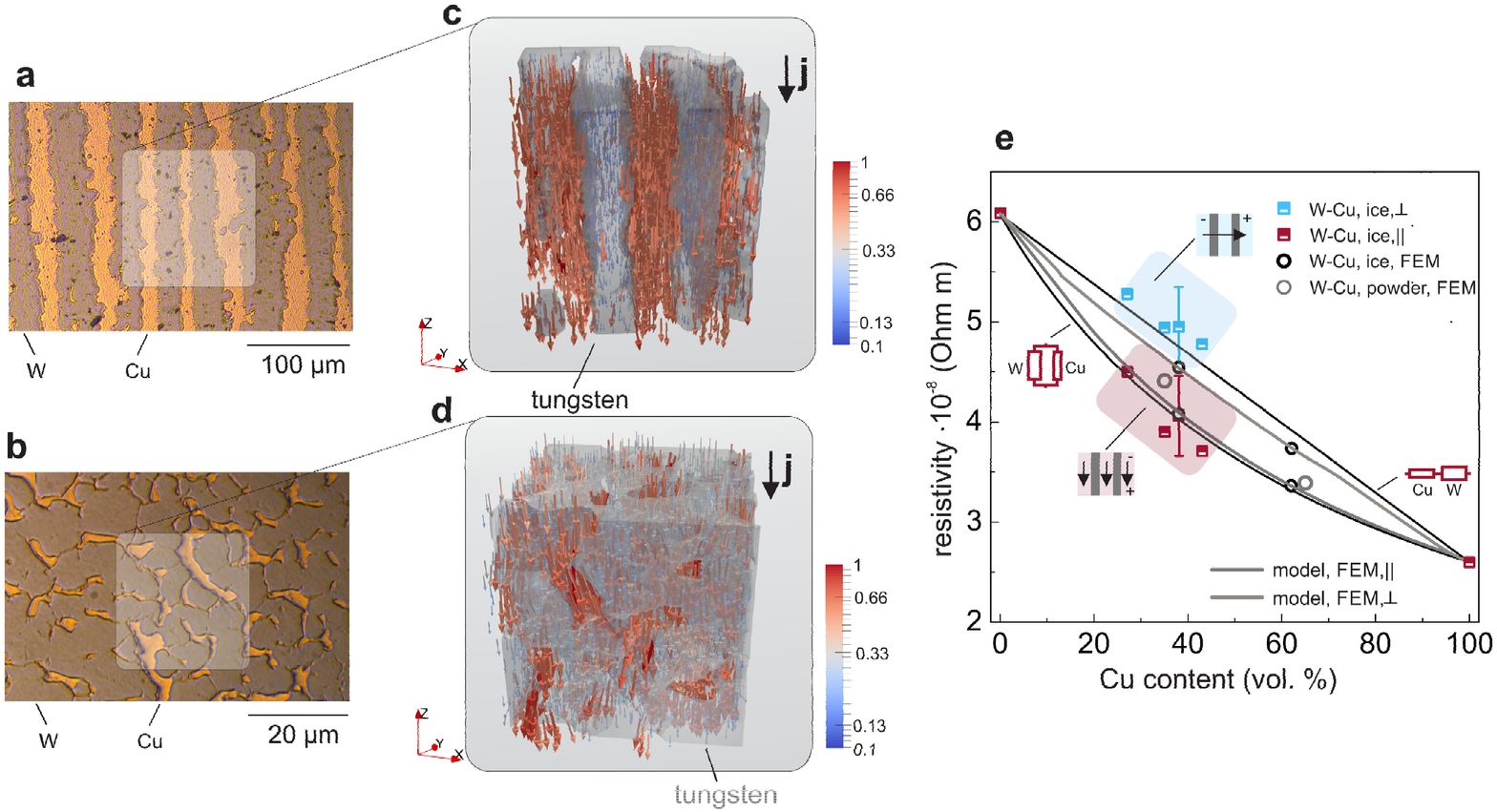}
\caption{\textbf{Electronic Transport.} Panel (a) and (b) show polished cross-sections of ice-templated and powder-metallurgy templated W-Cu acquired using light microscopy. The electronic transport properties are modelled using FEM on a subset of the measured XCT data. Panel (c) and (d) present the simulated local current flow (arrows) and normalized current density distribution (arrow colour) within an ice-templated (c) and powder-metallurgy (d) W-Cu composite. The high resistivity regions, i.e. the W-phase (grey), correlate with regions of low current flow. (d) Comparison of the anisotropic composite resistivity observed in experiments with theoretical predictions based on the FEM model illustrated in panel (c)-(d) and a parallel or linear combination of two resistors.}    
\label{fig:4}
\end{center}
\end{figure*}
%%%%%%%%%%%%%%%%%%%%%

Using the experimental three-dimensional XCT data, we can build representative FE models to study the electronic transport properties of the fabricated W-Cu composites. Of special interest is the impact of structural anisotropy on the composites resistivity. When current or heat is transported through a lamellar structure, one can imagine two distinct situations, namely the flow of current parallel to the lamellae and across the lamellae. In both cases the local current density is given by Ohm‘s law $\mathrm{j}(\mathrm{r})=\mathrm{\rho}(\mathrm{r})^{-1}\mathrm{E}(\mathrm{r})$, but the local resistivity contrast between the two phases, i.e. the composite architecture  molds the current flow within the structure. In our samples (Fig.\ref{fig:4}(c)-(d)), due to the resistivity contrast between the W ($6.0\pm0.4\cdot10^{-8}$~$\Omega$m) and Cu-phase ($2.6\pm0.4\cdot10^{-8}$~$\Omega$m), the current is predominately transported within the Cu phase as shown by FEM simulations for ice-templated and powder-based composites, respectively. In both composites, the current density within the Cu-phase is enhanced by up-to one order of magnitude, see (Fig.\ref{fig:4}(c)-(e)). To this extend, we can introduce an effective resistivity that treats the current transport in the composite within the bounds of two parallel resistors for the lower limit and two resistors in series for the upper limit. Thereby the resistivity of each resistor corresponds to the pure Cu and W phase, respectively. By applying experimental values for these two parameters, we can calculate the resistivity as function of Cu content and model the resistivity by solving Ohm's law within the FEM model. In Figure \ref{fig:4}(e) the experimentally determined resistivity values are compared with these theoretical predictions. The electrical resistivity of the composites, bulk copper, as used for the infiltration, and of bulk tungsten were measured by using the four-point probe method: for the pure metals values of $2.6\pm0.4\cdot10^{-8}$~$\Omega$m for Cu and $6.0\pm0.4\cdot10^{-8}$~$\Omega$m for W were found. The determined values agree with values from literature~\cite{Serway1998,Phillips2009}.
Using a linear array probe head enables a directional measurement of the resistivity parallel and perpendicular to the lamellae in the ice-templated structure. The corresponding resistivity values are reported and highlighted in Fig.\ref{fig:4}(e) by the light blue and pink areas.\\
The FEM simulations and the circuit models reproduce well the experimental results, confirming the anisotropic electronic transport properties of the lamellar composite. The lamellar structure of the composite, introduces a directional dependence of the current flow, which results in anisotropic electrical resistance as shown in Fig.\ref{fig:4}(e). As expected, the powder-based composite exhibits isotropic behaviour (Fig.\ref{fig:4}(e),light grey circles), while the overall resistivity exceeds the parallel resistor circuit. The FEM simulation indicate that the current transport in the isotropic composite is confined to the Cu-phase, see Fig.\ref{fig:4}(d), whose tortuosity controls the composite resistivity. A tortuous path effectively  increases the conductive path and hence the resistivity of the composite, as compared to the parallel resistor model without tortuosity.  \\
It is noteworthy, that similar effects, due to the analogy of Ohm's law and Fourier's law, can be expected for heat conduction given a significant difference in thermal conductivity of both materials~\cite{Boernstein1991}. 
\subsection*{Mechanical Properties}
%%%%%%%%%%%%%
\begin{figure*}[h!]
\begin{center}
\includegraphics[width=0.98\textwidth]{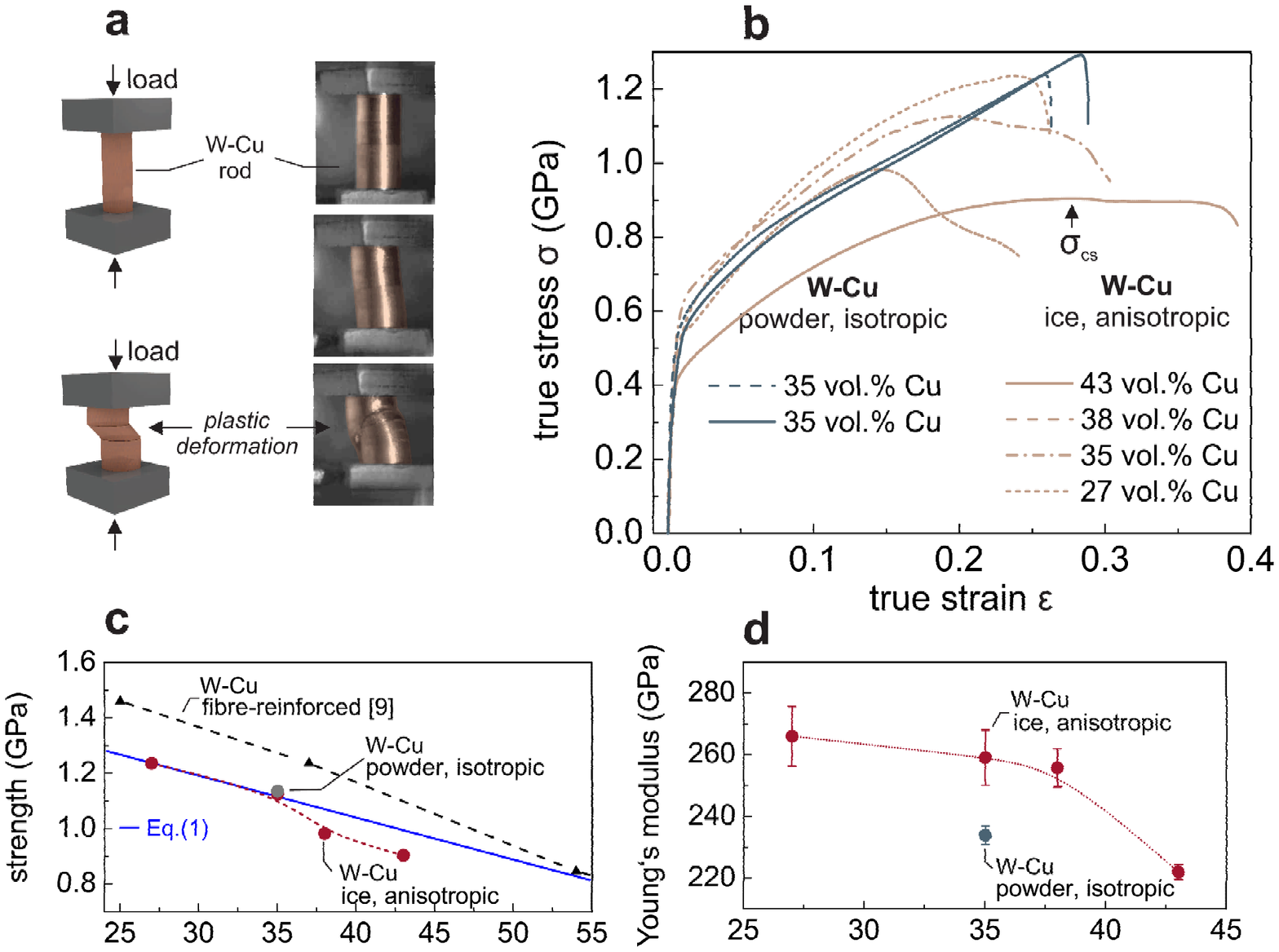}
\caption{\textbf{Mechanical behavior.} Compressive properties of ice-templated and powder-metallurgy W-Cu composites under compression. Panel (a) shows a schematic illustration and photographs of the compressive tests of an ice-templated W-Cu composite with 43 vol.$\%$ Cu. (b) Experimental stress-strain curves of W-Cu composites with various Cu volume fractions. All ice-templated composites, measured with lamellae parallel to the applied stress, feature a region of plastic instability after the yield strength is reached. The ultimate compressive strength %defined as the maximal stress state reach during compression 
is labelled $\sigma_{\text{cs}}$ as an example for a composite with 43 vol.$\%$ Cu. Panel (c) compares the measured compressive strength of (Fig.3)b with theoretical predictions based on a simple rule of mixture (Eq.~\ref{eq:1}) and compressive strength measured for fibre-reinforced W-Cu composite tested in the fiber direction, from Ref.[9]. (d) Dependence on Cu fraction of Young's moduli determined from stress-strain curves of W-Cu composites.}    
\label{fig:3}
\end{center}
\end{figure*}
%%%%%%%%%%%%%%%%%%%%%
The mechanical properties of the composites are determined by uni-axial compression test (Figure ~\ref{fig:3}a), which is an ideal technique to characterize the elastic and plastic response of a material to an external load. In Figure~\ref{fig:3}a the different stages of deformation - from elastic to plastic deformation to shearing are depicted by photographs taken during testing. 
The stress-strain curves in Figure~\ref{fig:3}b show distinct behavior depending on the composites' architecture. In both cases, no sharp yield point is visible but rather a gradual transition from the elastic to the plastic regime is observed. The powder-based isotropic composite exhibits a classical transition from linear elastic to linear plastic deformation terminated by brittle fracture once the compressive strength is reached. By contrast, the ice-templated composites feature a different mechanical response. The stress-strain curves of the anisotropic composite exhibits next to its linear elastic region a nonlinear plastic region including softening after the compressive strength of the material is reached. %\sout{In this softening region where $d\sigma/d\epsilon<0$, the material is unstable with respect to the Drucker criterion~\cite{Drucker1959} and the stress fields within the material are typically relaxed by local formation of cracks.} 
The measured mechanical properties, such as compressive strength and Young's modulus are summarized in Table~1.\\
%%%%%%%%%%%%%%%%%%%%%
\begin{table}
\label{tab:1}
    \begin{tabular}{llllll}
    \\
    ~ & $f_\text{Cu}$ [vol.\%]& $W$ [$\mu$m] & $E$ [MPa] &  $\sigma$ [MPa] & $\kappa$~[J/m$^3$ $10^4$ ]\\ \hline
    ice & $27$ & $49$ & $266$   & $1236$  & $308$\\ 
    ice & $35$ & $28$ & $259$   & $1125$  & $196$\\
    ice & $38$ & ~    & $255$   & $983$  & $294$\\
    ice & $43$ & ~    & $222$   & $904$  & $262$\\ \hline
    powder & $35$ & ~    & $234$   & $1136$  & $261$\\
    \end{tabular}
    \caption{Mechanical and structural properties including: volume fraction  $f_\text{Cu}$, lamella thickness $W$, Young’s modulus $E$, compressive strength $\sigma$ and toughness $\kappa$, i.e. stored energy density in compression}
\end{table}
%%%%%%%%%%%%%%%%%%%%%
The origin of strain softening can be explained by considering the interaction between the strong, brittle phase (W) and the weak, ductile phase (Cu) in the composite, whose yield stress inducing plastic flow is much smaller than the failure stress of the former. In such a situation, the plastically flowing Cu phase interacts with elastically deformed W-lamellae via shear forces at the Cu/W interface. For such a composite with strong fibers within plastically-deforming matrix, Kelly and Tyson~\cite{KELLY1965} have introduced a linear rule of mixture that captures these two coupled deformation mechanisms. This rule of mixture, suited for platelet-based composites as well~\cite{Glavinchevski1973}, superimposes the ultimate strength $\sigma_{us,i}$ of the single components to predict compressive strength $\sigma$,leading to: 
\begin{equation}
\sigma=\alpha\cdot\sigma_\text{us,W}(1-\phi)+\sigma_\text{us,Cu}\phi.
\label{eq:1}    
\end{equation}
Here, $\phi$ is the volume fraction of the Cu-phase and $\alpha=1-\frac{\sigma_{us,\text{W}} t}{\tau_{y,Cu} L}$ is the reinforcement factor containing the ratio between the ultimate strength of tungsten $\sigma_{us,W}$ and the shear strength of copper $\tau_{y,\text{Cu}}$ and the ration between the lamellae thickness $t$ and the lamella length $L$.\\
\\
We validate the model against the experimental results, using values $\sigma_\text{cs,W}=1686$~MPa, $\sigma_\text{cs,Cu}=147$~MPa. from Ref.23 and $\alpha=0.96$ with values for $\sigma_{us,W}$ and $\tau_{y,\text{Cu}}$ from Ref.\cite{schmidt1963engineering} and ~\cite{davis2001copper}. Figure \ref{fig:3}c reports the calculate compressive strength using Eq.~\ref{eq:1} and the experimentally measured compressive strength as function of Cu content, while Fig.\ref{fig:3}d shows the measured Young's modulus. The linear decrease in strength predicted by Eq.~\ref{eq:1} is well represented in our experimental data for both composites, although some deviation is found for higher Cu-contents. Good agreement for both composite types with the rule of mixture originate from the fact that the elastic and plastic deformation until the ultimate strength is reached seem only slightly affected by the composite structure. In the elastic regime, the tungsten skeleton is carrying a majority of the load, as copper yields below 50 MPa. Since tungsten is very brittle at room temperature, little to no plastic behavior of the tungsten skeleton is expected~\cite{Lassner1999}. Therefore the observed ductility of the composite until ultimate strength is attributed to the plastic deformation of the copper matrix, despite the presence of a continuous tungsten phase~\cite{Belk1993}. The compressive strength and Young's modulus rise with decreasing copper content (Figure~\ref{fig:3}c-d) in the ice-templated composites, as result from the much higher strength and stiffness of the tungsten phase. It is to note, that the scattering of the compressive stiffness data is attributed to measuring compressive stiffness during loading, as the composites were tested without load-unload loops.\\
The effect of anisotropy on the mechanical response of the composites comes into play once the ultimate strength ($\sigma_{cs}$) is reached. Beyond this critical value the mechanical response is nonlinear as featured in the stress-strain curves and exhibits significant softening, see Fig.~\ref{fig:3}b. In contrast to the powder-based composites, the lamellar architecture of the ice-templated W-Cu structures offer no direct path for cracks to propagate perpendicular to the load axis suppressing brittle fracture at the $\sigma_{cs}$. 
Rather, it is expected that the applied compressive stress will produce a local shear stress in regions where the W-lamellae are misaligned with respect to the compression direction. The initial shear and rotation in these regions is the onset for "kink band" formation, a deformed and rotated region within the compressed sample, which is a common deformation mode in anisotropic composites~\cite{argon2013fracture,MORAN19952943,BENABOU2010335}, such as Cu-Nb nanolaminates~\cite{nizolek2015kink,nizolek2017strain}. This deformation mode is expected in our lamellar structure with mechanical anisotropy under compression. The strain softening observed in Fig.~\ref{fig:3}b and formation of a geometrical deformed band during compression, as shown in the Supplementary Information Video S1 and SFig.3, are key characteristics for kink band formation. The formation of the kink band localizes the plastic deformation of the sample during compression as shown in Video 1, resulting in kink-band broadening~\cite{MORAN19952943} with increasing strain. 
%With a higher copper content, this shearing of the matrix is more pronounced as the volume fraction of the brittle tungsten is lower. Analogously, a higher tungsten content decreases the shear potential as is shown for the stress-strain curve of the low copper content composite (Fig.~\ref{fig:3}b). The presented results are in good agreement with kink band formation recently found in Cu-Nb .

%%%%%%%%%%%%%%% Conclusion
\section*{Conclusion}

We have demonstrated a template-based approach to fabricate anisotropic refractory metal-based composites. Here, ice-walls grown in a thermal gradient have been used to create lamellar W-Cu composite with a highly anisotropic lamellar architecture. Using this technique lamellar composite architectures or laminates with an average lamella spacing of $34$~$\mu$m and lamella thickness between $28-49$~$\mu$m have been produced. We have studied the 
composites' architecture including electronic and mechanical properties in direct comparison with a purely isotropic W-Cu composite. In particular, we observe a directional dependence of the composites resistivity, that can be engineered by changing the Cu content of the composite. Applying a combination of analytical and numerical techniques to the measured 3D XCT data, we demonstrate that the current density is enhanced by one order of magnitude within the Cu phase as compared to the W phase, while the overall composite resistivity can be explained in first approximation by a simple combination of two resistors in parallel or in series. Similar to fibre-reinforced composite, we demonstrate that strength of the lamellar composites follow a simple rule of mixture, whereby the anisotropy of the structure leads to nonlinear plastic behaviour and significant strain softening induced by kink band formation after the ultimate strength is reached.\\
The refractory metal based lamellar composites introduced  in  this  work open  up  new  possibilities to design anisotropic, mechanically enhanced materials especially for energy applications. Ice-templating is rapid, simple and potentially scalable to large areas, and can produce tailored anisotropic composites for a new generation of circuit breakers, thermal emitters and large scale photonic and plasmonic components. Here W-Cu is used as a testbed, but the introduced concept is applicable to a wide range of refractory metals and lower-melting, high-ductility matrices such as Cu alloys, Ni-alloys and Co-alloys. Furthermore, the novel infiltration process of the metal scaffold presented is not limited to ordinary metals, but can be easily extended to polymer-melts, ceramic-slurries with sintering temperatures below the melting point of the metal scaffold, and even liquid electrolytes.

\section*{Methods}

\subsection*{Composite Synthesis}

Aligned tungsten scaffolds were created via hydrogen reduction and sintering of a green body, created by ice-templating of an aqueous slurry of tungsten trioxide nanoparticles (WO$_3$, 99.95 $\%$ purity, $d < 100$~nm, SkySpring Nanomaterials Inc.) using 0.5 wt.$\%$  nickel oxide (NiO, 99.9$\%$ purity, $d = 20$~nm, Inframat Advanced Materials), 2.5 wt.$\%$  polyethylene glycol (PEG, Mn = 3,400, Sigma-Aldrich) and 1 wt.$\%$  polyethylene imine (PEI, Mw 25’000, Sigma-Aldrich) as sintering activator, binder and dispersant respectively. The process is schematically summarized in Figure~\ref{fig:1}. The effects of processing parameters - powder fraction in the slurry, freeze-casting and sintering temperatures as well as the influence of nickel - on the final scaffold microstructure, in particular open porosity and structural wavelength, have been reported in detail in a previous publication~\cite{Roethlisberger_2016}. The resulting anisotropic tungsten foam cylinders, were subsequently melt-infiltrated with Cu (99.99 $\%$  purity, Mateck GmbH) at $1300^\circ$C under Ar/5 $\%$  H$_2$ atmosphere (99.999 $\%$  purity, PanGas). The copper was placed on top of the tungsten scaffold in an alumina crucible (99.8 $\%$  purity Al$_2$O$_3$, 15 mm diameter, Metoxit GmbH) and melted by inductive heating. %To avoid thermal shocks due to rapid melting of copper, the inductive coupling was performed by wrapping the crucible mantle with conductive graphite tape.
The temperature was monitored optically by a custom-built  laser pyrometer. Upon melting, the copper infiltrates the tungsten foam due to gravitational forces. After solidification and cooling of the composites to room temperature, cylindrical compressive specimens (5-6 mm diameter and height/diameter aspect ratio of 2:1) were machined on a lathe. Control specimens with isotropic W distribution were obtained from W-20 wt.$\%$  Cu composites created via partial sintering of W powder preforms followed by Cu liquid infiltration (Plansee Powertech AG, Seon, Switzerland).

\subsection*{Microstructural, Electrical and Mechanical Characterization}

Radial and longitudinal cross-sections (perpendicular and parallel to the slurry solidification direction, respectively) were cut and prepared according to standard metallography preparation and imaged via light microscopy. To compare the ice-templated and isotropic W-Cu composite microstructures, high energy (acceleration voltage 170 kV) X-Ray tomography (Phoenix Nanotom S, GE Measurements) was performed (voxel size $1.3$~$\mu$m) on 400~$\mu$m diameter cylindrical composite samples machined with a lathe. The three-dimensional data where visualized using ImageJ and Blender, whereby Parallel FFTJ was been used to calculate the three dimensional fast Fourier transform (FFT) of the composites. To build the 3D FE-models the XCT-data have been used. Hereby, the 3D-surfaces of the different phases have been imported to Comsol Multiphysics\texttrademark. In case of lamellar composites, the volume fraction within the FE-model can been altered by anisotropically scaling the components of one of the two phases.
Compression tests were conducted at $25^\circ$C , well below the brittle-to-ductile-transition of tungsten reported~\cite{Lassner1999} as $280$--$330^\circ$C,  depending on strain rate~\cite{Giannattasio2007} and impurity level~\cite{Giannattasio2010}. In all cases, the load axis was parallel to the lamellar structure. A screw-driven mechanical testing machine operated at constant crosshead displacement rate (corresponding to a sample strain rate of ~$10^{-4}$s$^{-1}$) was used, with sample strain measured optically using digital image correlation.
Four-point resistivity measurements (National Instruments, NI PXI-4071 and Agilent, N6700B 400W, for voltage measurement and current supply, respectively) were conducted in steady-state conditions with current flow parallel and perpendicular the lamellar structure to obtain the anisotropic electrical resistivity of the composites. For this purpose, a Jandel Multiheight probe station equipped with a Jandel linear four point probe head (tip spacing$=1$mm) was used. These measurements were performed on polished disk-shaped samples with a typical diameter of 15 mm and height of 3 mm.

\section*{Acknowledgements }

The authors thank the Scientific Center for Optical and Electron Microscopy (ScopeM) at the ETH Zürich and the Optical Microscopy and Metallography Facility at Northwestern University (NU) for their support. We thank V. Wood from the Laboratory for Nanoelectronics, ETH Zurich for access to her high resolution x-ray computed tomography system. The project was funded by the Competence Center for Materials Science and Technology (CCMX), Switzerland. 
\section*{Author contributions statement}

A.R.  designed  and performed  the  experimental  research. A.R. and S.H.  fabricated  the  samples  used  in  the  article. A.R. performed and analyzed the mechanical measurements. H.G. performed and analyzed the simulations, the structural analysis and resistivity measurements. D.C.D and R.S suggested experiments and contributed to the interpretation.  All  authors  contributed  equally  to  the  preparation of the manuscript.

\section*{Additional information}

\textbf{Competing financial interests} The authors declare no competing financial interests.

%% Put the bibliography here, most people will use BiBTeX in
%% which case the environment below should be replaced with
%% the \bibliography{} command.

% \bibitem{dummy} Articles are restricted to 50 references, Letters
% to 30.
% \bibitem{dummyb} No compound references -- only one source per
% reference.
% \end{thebibliography}

%% Here is the endmatter stuff: Supplementary Info, etc.
%% Use \item's to separate, default label is "Acknowledgements"

%\bibliography{MLbibliography.bib}
%\noident

\end{document}